\documentclass[10pt, superscriptaddress,amsmath,amssymb,aps,prl,twocolumn,floatfix]{revtex4-2}
\usepackage{graphicx}
\usepackage{float}
\usepackage{bm}
\usepackage[colorlinks, linkcolor=mycol, citecolor=mycol, urlcolor=mycol, breaklinks]{hyperref}
\usepackage{braket}
\usepackage{bbold}
\usepackage{xcolor} 
\usepackage{comment}
\usepackage{physics}
\usepackage{verbatim}
\usepackage{upgreek}

\newcommand{\mi}{ {\rm i} }

\usepackage{amsmath}
\usepackage{amssymb}

\definecolor{mycol}{RGB}{10,55,130}

\begin{document}

\author{T.~Botzung}
\affiliation{CESQ/ISIS (UMR 7006), CNRS and Universit\'{e} de Strasbourg, 67000 Strasbourg, France}

\author{G.~Creutzer}
\affiliation{Laboratoire Kastler Brossel, Collège de France, CNRS, ENS-Université PSL, Sorbonne Université, 11 place Marcelin Berthelot, F-75231 Paris, France}

\author{C.~Sayrin}
\affiliation{Laboratoire Kastler Brossel, Collège de France, CNRS, ENS-Université PSL, Sorbonne Université, 11 place Marcelin Berthelot, F-75231 Paris, France}

\affiliation{Institut Universitaire de France, 1 rue Descartes, 75231 Paris Cedex 05, France}

\author{J.~Schachenmayer}
\affiliation{CESQ/ISIS (UMR 7006), CNRS and Universit\'{e} de Strasbourg, 67000 Strasbourg, France}

\begin{abstract}
    We introduce a method to synthetically engineer the range of dipolar interactions in tweezer atom arrays by effectively modifying the modes of the electromagnetic vacuum with far-detuned relay atoms. We derive equations of motion for the atoms of interest after adiabatic elimination of the relay atoms. We show the effectiveness of the scheme for realistic experimental parameter regimes with circular and low-angular-momentum Rydberg atom states.
\end{abstract}

\title{Tailoring interaction ranges in atom arrays}
\date{\today}
\maketitle

Rydberg-atom arrays are a powerful platform for quantum simulation and computation~\cite{Saffman2010, Browaeys2020Feb, Morgado2021, Wu2021}. Their ability to scale to large system sizes~\cite{Gyger2024Feb, Tao2024, Manetsch2024Mar, Lin2024Dec, Pichard2024}, maintain long coherence times~\cite{Barnes2022May, Graham2023, Huie2023, Norcia2023, Bluvstein2024Feb, Manetsch2024Mar}, and achieve reconfigurable lattice geometries~\cite{Kim2016Oct, Bernien2017Nov, Barredo2016,Bluvstein2022Apr, Bluvstein2024Feb} makes them  well-suited for studying complex quantum many-body phenomena. Analog quantum simulations~\cite{Bernien2017Nov, Keesling2019Apr, Ebadi2021Jul, Semeghini2021Dec, Browaeys2020} have provided insights into quantum magnetism~\cite{Bijnen2015, Zeiher2017, Leseleuc2018, Scholl2021Jul}, symmetry-protected topological phases~\cite{deLeseleuc2019Aug}, coherent excitation transfer ~\cite{Barredo2015Mar, Lienhard2020May, Han2024Mar, Li2025Mar}, emergent gauge fields~\cite{Zhang2018Nov, Celi2020Jun, Surace2020May, Cheng2024Sep}, and many-body localization~\cite{Marcuzzi2017Feb, Sahay2021Mar}. Circular Rydberg Atoms (CRA), i.e., states with maximal angular momentum, already used in cavity quantum electrodynamics~\cite{Haroche2013Jul} and quantum sensing~\cite{Facon2016Jul, Dietsche2019Apr}, are promising for quantum simulation and computation~\cite{Nguyen2018Feb, Meinert2020May, Cohen2021Aug} due to their long lifetimes~\cite{Cantat2020}.

Tailoring interactions is a challenge for analog quantum simulators. Ion-trap platforms have demonstrated variable-range interactions in the long-range regime, characterized by a power-law decay ($\sim1/r^{b}$) with exponents between $0\lesssim b \lesssim  3$~\cite{Britton2012Apr, schachenmayer2013entanglement, Grass2014Dec, Islam2013May, Richerme2013Sep, Kim2009Sep, Wang2013Jan}. In Rydberg atoms, exponents are generally fixed by near-field dipole-dipole couplings, $1/r^{3}$~\cite{Saffman2010, Walker2008Mar}, or to an effective van der Waals scaling of $1/r^{6}$. Recent advances have shown potential for  tunability, including modified pair interactions~\cite{Hummel2024Oct}, controlled excitation transport~\cite{Han2024Mar}, and enhanced superradiance in hybrid Rydberg-cavity systems~\cite{Han2024Dec}. 

\smallskip

In this article, we propose a method to engineer an effective variable-range spin exchange model, with an effective power-law behavior in the range $3 \lesssim b \lesssim 6$. We achieve this by modifying the electromagnetic environment with far-detuned relay atoms. Through adiabatic elimination of relay atoms, we demonstrate that the interaction can be finely tuned and depends solely on their position. We demonstrate the feasibility of the scheme with an array of CRAs~\cite{Ravon2023} with far-detuned low-angular-momentum Rydberg states as relays. 

\begin{figure*}[htb]
    \centering
    \includegraphics[width=1\linewidth]{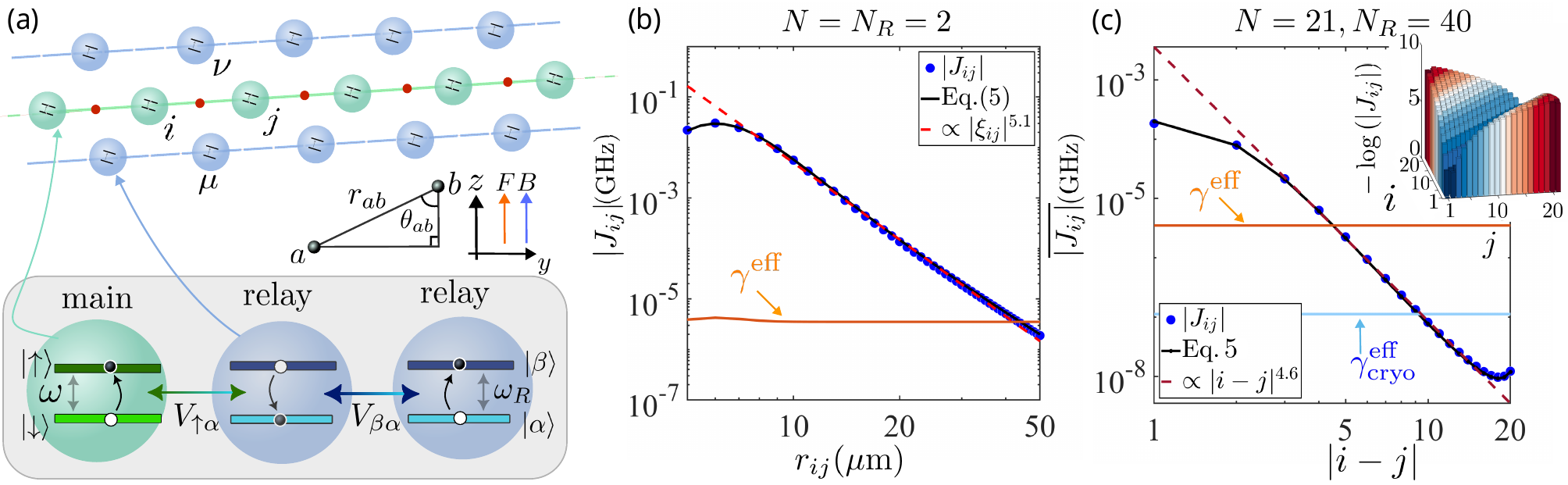}
    \caption{(a) Setup: $N$ main atoms (green, $i, j= 1,\dots,N$) are placed on a line, relay atoms (blue, $\mu,\nu = N+1,\dots,M$) are located symmetrically above and below (see text for detailed parametrization). We define the pairwise atom spacing $r_{ab}$, and angles $\theta_{ab}$ as sketched. A static electric (magnetic) field $F$ ($B$) points along the $z$-axis. Gray box: For main (relay) atoms we consider two-level systems with states $\ket{\downarrow}$, $\ket{\uparrow}$ ($\ket{\alpha}$, $\ket{\beta}$) and spacing $\omega$ ($\omega_R$). We include dipole-dipole couplings between all atoms: main-relay, relay-relay, however, main-main couplings vanish in a ``magic angle'' configuration with $\cos\theta_{ij} = \sqrt{1/3}$. (b/c) Two examples of effective distance dependence of $J_{ij}$ for a realization with CRAs. Blue points: effective model [Eq.~\eqref{eq: effectivecoupling}], solid lines: analytical results [Eq.~\eqref{eq: analyticsimplify}], dashed lines:  power-law fits.  Nearly horizontal lines: effective decay rates at room ($T=300$ K) and cryogenic ($T=4$ K) temperatures, $\gamma^{\rm eff}$ and $\gamma^{\rm eff}_{\rm cryo}$, respectively. (b) $|J_{ij}|$ between two main atoms as function of distance $r_{ij}$ ($r_{i\mu}=6 \, \upmu$m, $\theta_{i\mu}=0$). (c) Averaged coupling $\overline{|J_{ij}|}$ in a chain of $21$ main atoms as function of $|i-j|$ ($r_{i,i+1}=10 \,\upmu$m, $r_{i\mu}=12 \, \upmu$m, $\theta_{i\mu}=0$). Inset: Full coupling matrix, $-\log{(|J_{ij}|)}$.  Parameters:  $C_3^{(\uparrow\alpha)} = \pi \times 2.25$ GHz$\cdot\upmu$m$^3$,  $C_3^{(\beta\alpha)} = \pi \times 1.31$ GHz$\cdot\upmu$m$^3$,  $p_{\uparrow \downarrow}^{(2)} = -0.39$ MHz/(V/cm)$^2$,  $p_{\beta\alpha}^{(2)} = -57$ MHz/(V/cm)$^2$,  $F = 3.5$ V/cm.}
    \label{fig: model}
\end{figure*}

\smallskip

\textit{Setup---} We propose a synthetic implementation of the spin-1/2 model Hamiltonian 
\begin{align}
    \hat H^{\rm eff} = \sum_{i<j} J_{ij} (\hat \sigma_i^\dag \hat \sigma_j + \text{h.c.}),
    \label{eq: ham_XX}
\end{align}
with a tunable coupling matrix $J_{ij}$ between spins $i,j = 1,\dots,N$. Here $\hat \sigma_i \ket{\uparrow}_i = \ket{\downarrow}_i$ is the lowering operator for spin $i$. We consider all spins to be represented by two internal atomic states of $N$ ``main atoms'' and denote their transition frequency as $\omega$. In our scheme, the tunability of the coupling matrix is achieved by placing $N_R$ ``relay atoms'' in the vicinity of the main atoms, as sketched in Fig.~\ref{fig: model}(a). The total atom number is $M=N+N_R$. The relay atoms are modeled as two-level systems with states $\ket{\alpha}_\mu$ and $\ket{\beta}_\mu$ (as convention, we use the two indices $\mu,\nu$ for the range $\mu,\nu= N+1,\dots,M$) and transition frequency $\omega_R$. In a specific implementation below, we will use two consecutive circular Rydberg states as main atoms, and two low-angular-momentum Rydberg levels as relay atom levels. The position of an atom $a$ is given by the vector $\bm{r}_a$. The dipole-dipole coupling depends on atomic distances $r_{ab} = |\bm{r}_{a} - \bm{r}_{b}|$ and the relative orientation of the dipoles, here parametrized by the angles $\theta_{ab}$ between the connection vectors $\bm{r}_{a} - \bm{r}_{b}$ and the quantization ($z$) axis set by static fields [see Fig.~\ref{fig: model}(a)]. 

\smallskip

 We focus on a regime with  atomic distances much smaller than transition wavelengths, i.e.~near-field dipole-dipole interactions dominate. We consider dipole-allowed transitions between the states $\ket{\downarrow}_i \leftrightarrow \ket{\uparrow}_i$ and $\ket{\alpha}_{\mu} \leftrightarrow \ket{\beta}_\mu$, which lead to coherent energy transfer between all atoms by virtual photon-exchange. This is governed by the pairwise Hamiltonian $\hat H^{\rm{DD}}_{mn'} = \sum_{a<b} V_{(mn)(m'n')}(r_{ab}, \theta_{ab}) \big(\sigma_{mn}^{a} \sigma_{n'm'}^{b}  + \text{h.c.} \big)$ with $\sigma^{a}_{mn} = \ket{m}\bra{n}_a$~\cite{Lehmberg1970}, defined for all atoms, $a,b = 1,\dots,M$, and for the different internal state combinations, ${(mn)(m'n')} \in \{ (\uparrow\downarrow)(\uparrow\downarrow), (\uparrow\downarrow)(\beta\alpha), (\beta\alpha) (\beta\alpha) \}$. For conciseness, we shorten the notation $(mn)(m'n')$ to $(mn')$. In our convention, $V_{(mn')} (r_{ab}, \theta_{ab}) = C_3^{(mn')} (3\cos^{2}(\theta_{ab})-1)/\xi_{ab}^3$, with $\xi_{ab} = 2\pi r_{ab}/\lambda_{mn}$ for a transition wavelength $\lambda_{mn}$ and with constants $C_3^{(mn')}$ depending on the transition dipole moments.
 
\smallskip

Our scheme depends on the strength of a static electric field $F$ aligned along the $z$-axis, which modifies the bare Hamiltonian of each atom via a second-order Stark shift to $\hat{H}_{0} = \sum_{i} (\omega + p_{\uparrow}^{(2)} F^2) \ket{\uparrow}\bra{\uparrow}_i + \sum_{\mu} (\omega_R + p_{\beta}^{(2)} F^2) \ket{\beta}\bra{\beta}_\mu$ where $p_{\uparrow,\beta}^{(2)}$ are the differential second-order Stark shifts for the respective states. Note that we use a convention where the state $\otimes_i \ket{\downarrow}_i \otimes_\mu \ket{\alpha}_\mu$ corresponds to zero energy. The electric field $F$ provides control over the detuning from a Förster resonance condition~\cite{Anderson1998, Mourachko1998, Anderson2002, Anderson2005} between the main and the relay states, $\Delta = (\omega_R + p_{\beta\alpha}^{(2)}F^2) - (\omega + p_{\uparrow\downarrow}^{(2)}F^2)$. Summarizing all Hamiltonian contributions, the dynamics of the $M$ atoms is governed by $\hat H_M = \hat H_0 + \hat H_{\uparrow\downarrow}^{\rm DD} + \hat H_{\uparrow\alpha}^{\rm DD} + \hat H_{\alpha\beta}^{\rm DD}$. 

\smallskip

In our model, dissipation plays an important role. We consider full dynamics of the density operator $\rho$ governed by the Lindblad master equation ($\hbar \equiv 1$)
\begin{equation}
\label{eq: full_quantum_master}
\dot \rho = -\mi [\hat H_M, \rho] + \sum_\eta \mathcal{L}_\eta\rho \,
\end{equation}
with the superoperator for a noise channel $\eta$ defined as $\mathcal{L}_\eta \rho \equiv [\hat L_\eta \rho, \hat L_\eta^\dag] + [\hat L_\eta, \rho \hat L_\eta^\dag]$ with Lindblad jump operators $\hat L_\eta$. For the experimental realization described below, we include dissipation due to thermally induced re-population between the high-lying Rydberg states (black-body radiation, BBR), and spontaneous emission. The corresponding Lindblad operators and respective rates are defined as $\hat L_{mn} \equiv \sqrt{\gamma_{mn}} \dyad{m}{n}$. Note that in our implementation, the energy spacing between all states for which BBR is relevant is very similar, and we consider only a single BBR rate $\gamma\approx\gamma_{mn}$ for all $mn$ combinations.

\smallskip

The geometric configuration of our proposed setup is sketched in Fig.~\ref{fig: model}(a): We consider all main atoms placed on a line, such that $\theta_{ij} = \arccos(\sqrt{1/3})$ for $i,j = 1,\dots,N$. In such a ``magic angle'' configuration direct dipole-dipole coupling terms (and higher order van-der-Waals interactions induced between the atomic levels) vanish, and  $\hat H_{\uparrow\downarrow}^{\rm DD} = 0$. Still, effective interactions of the form of Hamiltonian~\eqref{eq: ham_XX} are induced via virtual excitations of the relay atoms. The goal of this article is to show how a controlled placement of the relay atoms leads to a tunable coupling matrix $J_{ij}$.
    
\medskip

\textit{Adiabatic elimination---}  We derive an effective equation of motion for the sub-space density matrix of the $N$ main atoms $\rho_N$, using adiabatic elimination~\cite{Zwanzig1960, Reiter2012, Schutz2020, hagenmuller2020}. Following the procedure in~\cite{hagenmuller2020}, the evolution on the Hilbert sub-space of the relay atoms is governed by the equation $\dot \rho = -\mi \hat{\bm{\tau}}^T \bm{M} \hat{\bm{\tau}} \rho + \mi \rho \hat{\bm{\tau}}^T \bm{M}^* \hat{\bm{\tau}} + 2 \gamma \hat{\bm{\tau}}^T  \rho \hat{\bm{\tau}}$. Here we defined the vector of relay lowering operators $\hat{\bm{\tau}}$ with elements $(\hat{\bm{\tau}})_\mu = \ket{\beta}\bra{\alpha}_\mu$. The complex $N_R\times N_R$  matrix $\bm{M}$ is given by the elements $(\bm{M})_{\mu\nu} = (\Delta - \mi \gamma)\delta_{\mu \nu} + (1 - \delta_{\mu\nu}) V_{\alpha\beta}(r_{\mu \nu}, \theta_{\mu\nu})$. Validity of the adiabatic elimination requires the $N_R$ eigenvalues $\bm{M}$ to be the largest energy scale in the system~\cite{hagenmuller2020}. This condition ensures that the relay atom dynamics takes place on a much faster timescale than the main atom dynamics~\cite{SOM}. Then, the evolution of the density operator projected on the Hilbert space of the main atoms is given by
\begin{align}
\label{eq: effective_master_equation}
    \dot\rho_N  = -\mi[\hat{H}^{\rm eff} + \hat{H}_0^{\rm eff}, \rho_N] + {\mathcal{L}}^{\rm eff} \rho_N,
\end{align}
where both the coherent and dissipative dynamics are now governed by effective  Hamiltonians and Lindblad superoperators $\hat{H}^{\rm eff}$, $\hat{H}_0^{\rm eff}$, and ${\mathcal{L}}^{\rm eff}$, respectively.

\smallskip

The effective Hamiltonian takes the form of Eq.~\eqref{eq: ham_XX} with modified couplings
\begin{align}
\label{eq: effectivecoupling}
\begin{array}{ccc}
    J_{ij} & = 
    - \text{Re}[\bm{V}_{j}^{\text{T}} \bm{M}^{-1} \bm{V}_{i}].
\end{array}
\end{align}
 Here, we defined vectors with $N_R$ elements corresponding to the couplings between the main atom $i$ and relay atoms $\mu$, $(\bm{V}_{i})_{\mu} = V_{\uparrow\alpha}(r_{i\mu}, \theta_{i\mu})$. Expression~\eqref{eq: effectivecoupling} is a consequence of second-order perturbation theory, where dynamics for the main atoms are modified by virtual dynamics in the ensemble of relay atoms. This is also seen more directly when neglecting the interactions between relay atoms. Then, Eq.~\eqref{eq: effectivecoupling} simplifies to
\begin{align}
\label{eq: analyticsimplify}
J_{ij} = \frac{1}{\Delta} \sum_{\mu} V_{\uparrow\alpha}(r_{i\mu}, \theta_{i\mu}) V_{\uparrow\alpha}(r_{j\mu}, \theta_{j\mu}).
\end{align}

\begin{figure*}[htb]
    \centering
    \includegraphics[width=1\linewidth]{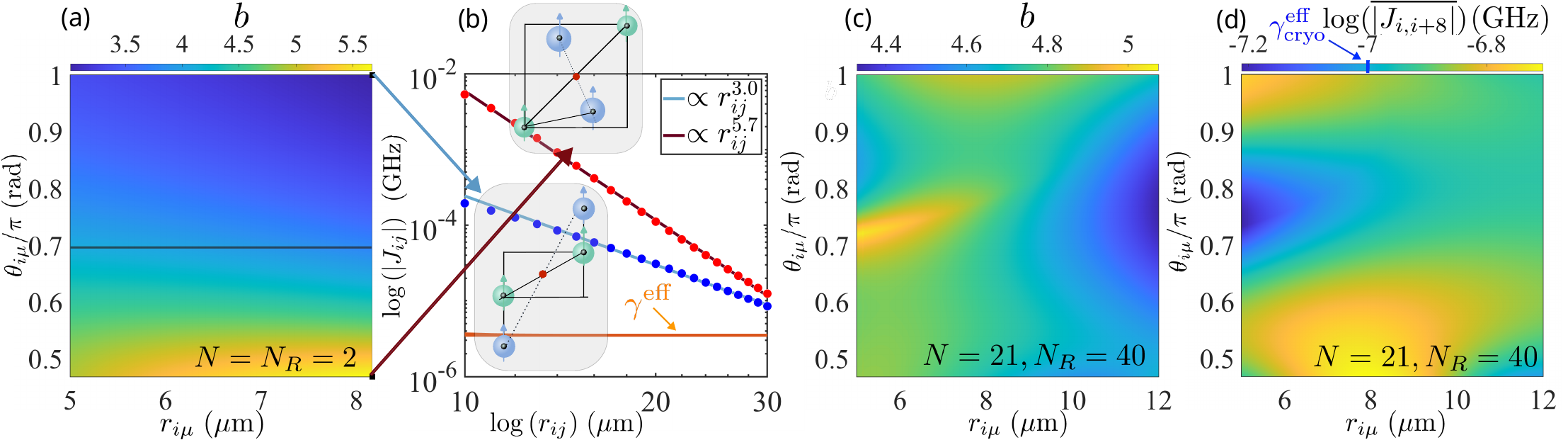}
    \caption{(a),(c) Contour plots of the power-law exponent $ b $ as a function of $r_{i\mu} $ and $\theta_{i\mu}$. It is obtained by fitting $ |J_{ij}| = a/r_{ij}^b $ and $ |J_{ij}| = a/|i-j|^b $ for panels (a), (c), respectively. (a) Pair of atoms, $N=N_R=2$. (c) Chain configuration $N=21, N_R=40$. (b) Examples of the distance dependence of $ |J_{ij}| $ at two specific configurations marked in (a). Insets: Corresponding geometrical configurations. (d) The averaged coupling strength over a distance of 8 atoms, $\log{(\overline{|J_{i, i+8}|)}}$. Parameters  identical to Fig.~\ref{fig: model}.}
    \label{fig: fig2}
\end{figure*}

We numerically confirm that neglecting relay atom interactions is an excellent approximation for our configuration in Fig.~\ref{fig: model}(b/c) and Fig.~\ref{fig: fig2}(b). From expression \eqref{eq: analyticsimplify} we observe that the effective pairwise interaction can be tuned in terms of distance, using the relay atom separations $r_{i\mu}$, $r_{j\nu}$, and in terms of sign, using the polar relay atom position angles $\theta_{i\mu}$, $\theta_{j\nu}$. Note that to compute the effective interaction strengths and ranges below, we numerically evaluate the full expression~\eqref{eq: effectivecoupling}.

 The effective Hamiltonian~\eqref{eq: effective_master_equation} also includes modified single-atom terms $\hat{H}_0^{\rm eff} = \sum_i \delta^{\rm eff} \ket{\uparrow}\bra{\uparrow}_i$ with detunings $\delta^{\rm eff}  =  (\omega + p_{\uparrow}^{(2)}F^{2}) -\text{Re}[\bm{V}_{i}^{\rm T} \bm{M}^{-1} \bm{V}_{i}]$, which can be compensated by choosing an appropriate geometry~\cite{SOM}. 
The local dissipative rates in Lindblad operators for the main atoms modify to $\gamma^{\rm{eff}} = \gamma + \text{Im}[\bm{V}_{i}^{\rm T} \bm{M}^{-1} \bm{V}_{i}]$. We note that in the near-field regime, coherent interactions dominate, thus leaving the dissipative rates nearly unchanged. 

\smallskip

\textit{Experimental implementation---} Our proposal builds on a novel experimental platform that enables the preparation of arrays of interacting laser-trapped CRAs~\cite{Ravon2023, Mehaignerie2025}. This platform uses circular Rydberg levels of Rubidium but other platforms have also prepared~\cite{Teixeira2020} and trapped~\cite{Holzl2024} circular Rydberg levels of Strontium. With $n\sim50$, the lifetimes of these levels reach tens of milliseconds in a cryogenic environment~\cite{Cantat-Moltrecht2020}. The lifetime of the circular Rydberg level with principal quantum number $n$, denoted $\ket{n\mathrm{C}}$, is solely limited by the microwave (MW) transition to the lower-lying $\ket{(n-1)\mathrm{C}}$ state. At room temperature, MW BBR reduces the lifetime of CRAs to a few $100~\upmu\mathrm{s}$, with the strongest contributions stemming from stimulated emission and absorption on the $\ket{n\mathrm{C}}\to\ket{(n\pm1)\mathrm{C}}$ transitions. 

Quantum non-demolition measurements and local manipulations of circular Rydberg states have recently been performed through the interaction of the CRAs with relay atoms transiently excited to a low-$l$ Rydberg level~\cite{Machu2025}. This experiment uses a Stark-tuned Förster resonance between the $\ket{53\mathrm{C}}\to\ket{54\mathrm{C}}$ and $\ket{45S_{1/2},m_J=1/2}\to\ket{45P_{3/2}, m_J=3/2}$ transitions. We propose to implement our method by mapping the states of the main atoms and relay atoms onto the same four levels: $\{\downarrow, \uparrow\} = \left\{53\mathrm{C}, 54\mathrm{C}\right\}$ and $\{\alpha, \beta\} = \left\{45S, 45P\right\}$. The Förster resonance ($\Delta=0$) is reached at an electric field of $F=1.6~\mathrm{V}/\mathrm{cm}$; with $F=3.5~\mathrm{V}/\mathrm{cm}$, $\Delta = 550~\mathrm{MHz}$. 

To describe the relaxation of the CRAs, we add two neighboring circular Rydberg levels $\ket{\Uparrow} = \ket{55\mathrm{C}}$ and $\ket{\Downarrow} = \ket{52\mathrm{C}}$ to our model. The main atoms, now being 4-level systems, relax via Lindblad jump operators of the form $\hat{L}_{mn} = \sqrt{\gamma_{mn}} \dyad{m}{n}$, with $mn \in \{\Uparrow\uparrow, \uparrow\Uparrow, \uparrow\downarrow, \downarrow\uparrow, \downarrow\Downarrow, \Downarrow\downarrow\}$. At room temperature, all $\gamma_{mn}$ rates are similar and we consider a single BBR rate $\gamma\approx 3.5\,\mathrm{kHz}$. At $T=0\,\mathrm{K}$, only spontaneous emission must be considered, so $\gamma_{\downarrow\Downarrow} = \gamma_{\uparrow\downarrow} = \gamma_{\Uparrow,\uparrow} = 0$, and for simplicity we choose the natural decay rate of $\ket{54\mathrm{C}}$ for the other rates to be $\gamma_{\Downarrow\downarrow} = \gamma_{\downarrow\uparrow} = \gamma_{\uparrow \Uparrow} \equiv \gamma^\mathrm{cryo} = 24\,\mathrm{Hz}$.
 
The low-$\ell$ Rydberg levels, $\ket{45S}$ and $\ket{45P}$, decay towards the ground state or other Rydberg levels by the emission of optical or MW photons, respectively. At room temperature, the decay rates associated to these two processes are almost equal, with a common value $\gamma_\ell \approx 10\,\mathrm{kHz}$. For simplicity, we model the decay on the MW transitions with the Lindblad operators $\hat{L}_{mn} = \sqrt{\gamma_\ell} \dyad{m}{n}$, with $mn \in \{\alpha\beta, \beta\alpha\}$. The decay on MW transitions is, however, strongly reduced at cryogenic temperatures. Only the decay towards the ground state is relevant in this case. As analyzed in~\cite{SOM}, including additional losses of relay atoms shortens the timescale for observing modified interactions. Moreover, repumping into the desired low-$\ell$ state is, in principle, feasible without adversely affecting the interaction measurement.

Note that one could also encode the relay atom levels $\ket{\alpha}$ and $\ket{\beta}$ onto long-lived circular Rydberg levels instead of low-$\ell$ Rydberg levels at the expense of a slight increase of $\Delta$. For instance, $\Delta=1.5\,\mathrm{GHz}$ with $\{\downarrow, \uparrow, \alpha, \beta\} = \{70\mathrm{C}, 71\mathrm{C}, 72\mathrm{C}, 73\mathrm{C}\}$.
This would have the advantage of effectively suppressing decay to the ground state--which occurs only for low-$\ell$ states--leaving dephasing as a main decoherence effect~\cite{SOM}.

\smallskip

\textit{A pair of atoms---} Starting with the simple case of $N=2$, we show an example on how relay atom placement modifies the effective coupling in Fig.~\ref{fig: model}(b). Here, we choose a geometry with $N_R=2$ as sketched in Fig.~\ref{fig: fig2}(b) (insets): We parametrize the position of the relay atom $\mu$ using only its distance $r_{i \mu}$ from the main atom $i$, and the polar angle $\theta_{i\mu}$. The second relay atom $\nu$ is located in a mirrored position with respect to the main atom $j$. This geometry ensures resonance between the levels of the main atoms, as the relay atoms induce inverse shifts~\cite{SOM}. It also ensures that the sign of $J_{ij}$ only depends on the  detuning $\Delta$. In the example in Fig.~\ref{fig: model}(b) we use $r_{i\mu}=6 \, \upmu$m, $\theta_{i\mu}=0$ and find that the long-distance behavior of $J_{ij}$ decays as a power-law with exponent $b \sim 5.1$. Note that the price for tunability of the interaction range via virtual relay excitation is a reduction in effective coupling strength. However, long-range couplings remain significantly above the effective state lifetimes at room temperature, indicated by the nearly horizontal line in Fig.~\ref{fig: model}(b).

To obtain an analytical understanding, using the parametrization for the relay-atom positions and Eq.~\eqref{eq: analyticsimplify}, we can express the coupling-strength dependence as a function of $r_{i\mu}$ and $\theta_{i\mu}$ only. Using a Taylor expansion in $r_{ij}^{-1}$ one finds~\cite{SOM}
\begin{align}
    J_{ij} &= \frac{2}{\Delta}V_{\uparrow\alpha}(r_{i\mu},\theta_{i\mu}) \sum_{\alpha \geq 4} 
  \frac{B_\alpha (\theta_{ij}, r_{i\mu})}{r_{ij}^\alpha}.
  \label{eq: Jijtaylor}
\end{align}
The two leading-order coefficients are given by $B_4 = r_{i\mu} \cos(\theta_{ij} - \theta_{i\mu}) - 3r_{i\mu} \cos(\theta_{i\mu} + \theta_{ij})$, $B_5 = \frac{33}{2} r_{i\mu}^2 \cos(2\theta_{i\mu}) +15  r_{i\mu}^2 \cos(2\theta_{ij}) + 5 r_{i\mu}^2 \cos^2(\theta_{ij} - \theta_{i\mu})$. Further analytical expressions for $J_{ij}$ are provided in~\cite{SOM}. For example, in the limit $r_{ij} \gg r_{i\mu}$, we see that in leading order $J_{ij}$ exhibits a power-law decay $\sim r_{ij}^{-4}$.

We analyze the tunability of the distance dependence in Fig.~\ref{fig: fig2}(a/b) as function of $r_{i\mu}$ and $\theta_{i\mu}$. Depending on the placement of the relay atoms, we find a wide tunability of the effective exponent in the range $3 \lesssim b \lesssim 6$, which we extract from fits of the form  $ |J_{ij}| \equiv a/r_{ij}^{b}$. Here, we focus on relay-atom distances $r_{i\mu}$ of at least  $5\,\upmu$m, for which we have confirmed that the adiabatic elimination remains valid~\cite{SOM}. The horizontal line indicates the location where $\theta_{i\mu} = \pi - \arccos(\sqrt{1/3})$, a point at which the direct dipole-dipole coupling between the relay atom and its neighboring main atom vanishes. We show that around $\theta_{i\mu} = \pi - \arccos(\sqrt{1/3}) \pm 0.05$, the effective interaction is too weak compared to the decay rate $\gamma^{\text{eff}}$ in~\cite{SOM}. This exclusion region follows from the condition $J_{ij} < \gamma^{\text{eff}}$ at $r_{ij} = 25 \; \upmu\text{m}$.  We generally observe that the tunable power-law behavior should be resolvable experimentally over a wide parameter range, even at room temperature as effective couplings remain large compared to effective decay rates, $|J_{ij}| \gg \gamma^{\rm eff}$.

\smallskip

\textit{Chain of atoms---} For analyzing effective couplings in the chains configuration, we choose equally spaced main atoms with distances $r_{i,i+1} = 10\,\upmu$m. We consider a succession of blocks of two main and two relay atoms in the same mirrored position as for $N=N_R=2$ [see sketch in Fig.~\ref{fig: model}(a)]. Therefore, the relay-atom distance $r_{i\mu}$ and relative relay-atom angle $\theta_{i\mu}$ remain the only free parameters and are identical for all $i=1,\dots,N$. 

Fig.~\ref{fig: model} (c) shows the averaged magnitude $\overline{|J_{ij}|}$ as a function of the distance $|i-j|$ obtained from Eq.~\eqref{eq: effectivecoupling} (blue points). We choose $r_{i\mu}=12$ $\upmu$m and $\theta=\pi/2$. The black solid line corresponds to the expression neglecting relay-relay interactions, Eq.~\eqref{eq: analyticsimplify}, and is in good agreement with Eq.~\eqref{eq: effectivecoupling}. The dashed line represents a power-law fit of the form $|J_{ij}|=a/|i-j|^{b}$. We find that the power-law model remains a good approximation at long distances throughout our parameter regime~\cite{SOM}. At a distance of $\sim 4$ atoms, the effective coupling strength becomes smaller than the effective decay rate at room temperature, $\gamma^{\rm eff}$. Therefore, observing long-range power-law interactions will require cryogenic temperatures, leading to the effective decay rate $\gamma^{\rm eff}_{\rm cryo}$ (blue horizontal line).  

In Fig.~\ref{fig: fig2}(c) we analyze the tunability of $b$ in the chain, as a function of $r_{i\mu}$ and $\theta_{i\mu}$. We demonstrate a smooth tunability in the range $b \sim 4-5$. The shortest range coupling is achieved for relay atoms placed close to the main atoms (small $r_{i\mu}$), where we find a relatively large variability of $b$ as function of $\theta_{i\mu}$. However, for $0.65 \pi \lesssim \theta_{i\mu} \lesssim 0.85\pi$ nearest-neighbor main-relay couplings vanish, leading to very small $|J_{ij}|$, as analyzed in Fig.~\ref{fig: fig2}(d). There, we plot the averaged coupling over a distance of $8$ atoms. To assess whether the effective power-law coupling is observable, we compare it to the effective main atom decay rate at cryogenic temperatures. For configurations with $\theta_{i\mu} \sim \pi/2$, we find strong effective long-range power-law couplings with variable exponents as function of $r_{i\mu}$. For large relay atom distances $r_{i\mu} \gg r_{ij}$ we recover the $b\sim 4$ power-law behavior, as perturbatively expected from Eq.~\eqref{eq: Jijtaylor}.

\smallskip

\textit{Conclusion---} We have introduced a scheme to  engineer tunable dipolar interactions between Rydberg atoms in tweezer arrays, by modifying the electromagnetic vacuum with a proper positioning of relay atoms. Our approach enables control over the interaction range in the near-field regime, achieving effective power-law behavior at long distances with exponents tunable between 3 and 6. This tunability opens promising avenues for analog quantum simulation, especially when implemented with CRAs in cryogenic environments. Future work will explore experimental realizations, more complex geometries, e.g.~with sign-alternating couplings, and potential applications in quantum gate operations using relay-mediated dynamics.

\medskip

\textit{Acknowledgements---} We thank M.~Brune for very helpful discussions. T.B.~and J.S.~have been supported by the ERC Consolidator project MATHLOCCA (Grant nr.~101170485), by the CNRS through the EMERGENCE@INC2024 project DINOPARC, and by the French National Research Agency under the Investments of the Future Program project ANR-21-ESRE-0032 (aQCess), and the France 2030 program ANR-23-PETQ-0002 (PEPR project QUTISYM). Computations  were  carried  out  using  resources  of  the High Performance Computing Center of the University of Strasbourg, funded by Equip@Meso (as part of the Investments for the Future Program) and CPER Alsacalcul/Big Data. G.C.~and C.S.~have been supported by the Quantum Information Center Sorbonne as part of the program investissements d’excellence – IDEX of the Alliance Sorbonne Université. It has received funding by the France 2030 programs of the French National Research Agency (Grant No. ANR-22-PETQ-0004, project QuBitAF), under Horizon Europe programme HORIZON-CL42022-QUANTUM-02-SGA via the project 101113690 (PASQuanS2.1).
\bibliographystyle{apsrev4-1}
\bibliography{main}

\clearpage

\begin{widetext}

\setcounter{equation}{0}
\setcounter{figure}{0}
\setcounter{table}{0}
\setcounter{page}{1}
\renewcommand{\theequation}{S\arabic{equation}}
\renewcommand{\thefigure}{S\arabic{figure}}

\appendix

\def\appendixname{}
\renewcommand{\thesection}{\Roman{section}}
\setcounter{secnumdepth}{1}

\makeatletter
\def\@seccntformat#1{\csname the#1\endcsname.\quad}

\renewcommand\section{\@startsection{section}{1}{\z@}%
  {-3.5ex \@plus -1ex \@minus -.2ex}%
  {2.3ex \@plus.2ex}%
  {\centering\normalfont\normalsize\bfseries}}
\makeatother

\begin{center}

{\large{ {\bf Supplemental Material for: Tailoring interaction ranges in atom arrays\\ }}}

\vskip0.5\baselineskip{T.~ Botzung$^{1}$, G.~Creutzer$^{2}$, C.~Sayrin$^{2, 3}$, J.Schachenmayer$^{1}$}

\vskip0.5\baselineskip{ {\it 
$^1$CESQ/ISIS (UMR 7006), CNRS and Universit\'{e} de Strasbourg, 67000 Strasbourg, France\\
$^{2}$Laboratoire Kastler Brossel, Collège de France, CNRS, ENS-Université PSL, Sorbonne Université, 11 place Marcelin Berthelot, F-75231 Paris, France\\
$^{2}$Institut Universitaire de France, 1 rue Descartes, 75231 Paris Cedex 05, France
}}

\end{center}

\section{Robustness against dephasing}

\begin{figure}[htb]
    \centering
    \includegraphics[width=1\linewidth]{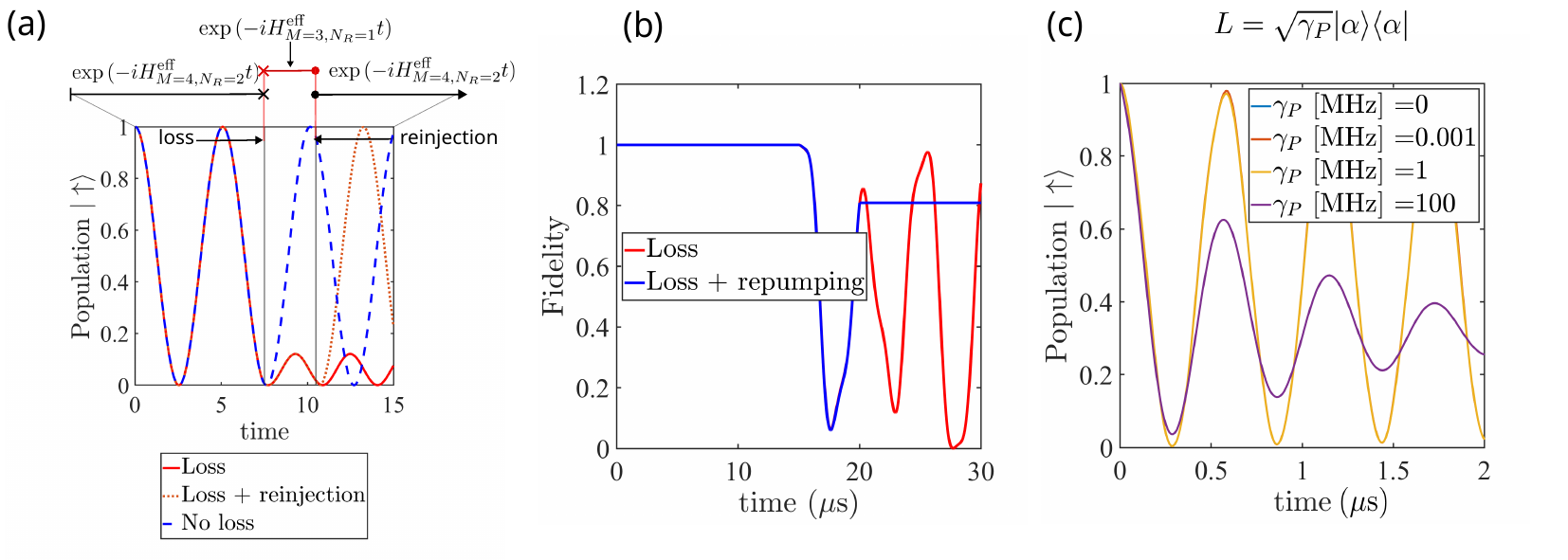}
    \caption{(a) Time evolution of the population in state $\ket{\uparrow}$ as predicted by the effective Hamiltonian dynamics for $N = N_R = 2$ [$r_{ij} = 15 \; \upmu$m, $\textbf{r}_{i\mu} = (6, 0)$]. Three scenarios are shown: evolution without relay atom loss (blue dots), with relay atom loss (red line), and with relay atom loss followed by repumping (orange dotted line).
    (b) Fidelity with respect to the no-loss evolution, compared to the cases with atom loss (red line) and with loss plus repumping (blue line).
    (c) Full master equation simulation of the $\ket{\uparrow}$ population dynamics including an additional jump operator $L = \sqrt{\gamma_P} \ket{\alpha} \bra{\alpha}$.}
    \label{fig: lossydynamics}
\end{figure}

In this section, we investigate how the loss of a relay atom affects the system's dynamics. This is motivated by the possible decay of the state $\ket{\alpha} = \ket{45S}$ to the atomic ground state, which limits the protocol described in the main text through the finite lifetime of $\ket{\alpha}$. We aim to understand how this loss impacts the dynamics and whether repumping into $\ket{\alpha}$ can mitigate its effects.

To this end, we propose two protocols. First, we consider an ideal regime without dissipation, where the adiabatic condition is satisfied, allowing us to use the effective model. We evolve the system with $H^{\rm eff}({M=4, N_R=2})$ until time $t_{\rm loss}$, at which point a relay atom is removed. The evolution then continues with $H^{\rm eff}({M=3, N_R=1})$ until $t_{\rm reinjection}$, after which we return to the initial Hamiltonian $H^{\rm eff}({M=4, N_R=2})$. This protocol is illustrated in Fig.~\ref{fig: lossydynamics}(a).

We compare three scenarios: no loss, loss without repumping, and loss followed by repumping into $\ket{\alpha}$. The results show that losing a relay atom suppresses energy exchange between the two main atoms, but repumping restores the exchange process as illustrated in Fig.~\ref{fig: lossydynamics}(a). However, looking at the fidelity of the state [Fig.~\ref{fig: lossydynamics}(b)] we see that the repumping protocol does not allow to continously simulate the quantum system. One can still infer the specific interaction scales, but usage as e.g. an analog quantum simulator will be limited by the finite lifetime of $\ket{\alpha}$. 

Secondly, we study the dynamics of the full open quantum model adding a dissipation of the form
\begin{equation}
    L =  \sqrt{\gamma_P} \ket{\alpha}\bra{\alpha}.
\end{equation}
In Fig.~\ref{fig: lossydynamics}(c) we show examples of the dynamics of the population of $\ket{\uparrow}$ for various rate $\gamma_P = 0, 0.001, 1, 100$ [MHz] (see colors in legend). We see that to observe significant effect, the rate $\gamma_P$ needs to be large indicating a strong robustness against the decoherence effect.

\section{Validity of the adiabatic elimination}
\label{ap : conditionadia}

\begin{figure}[htb]
    \centering
    \includegraphics[width=1\linewidth]{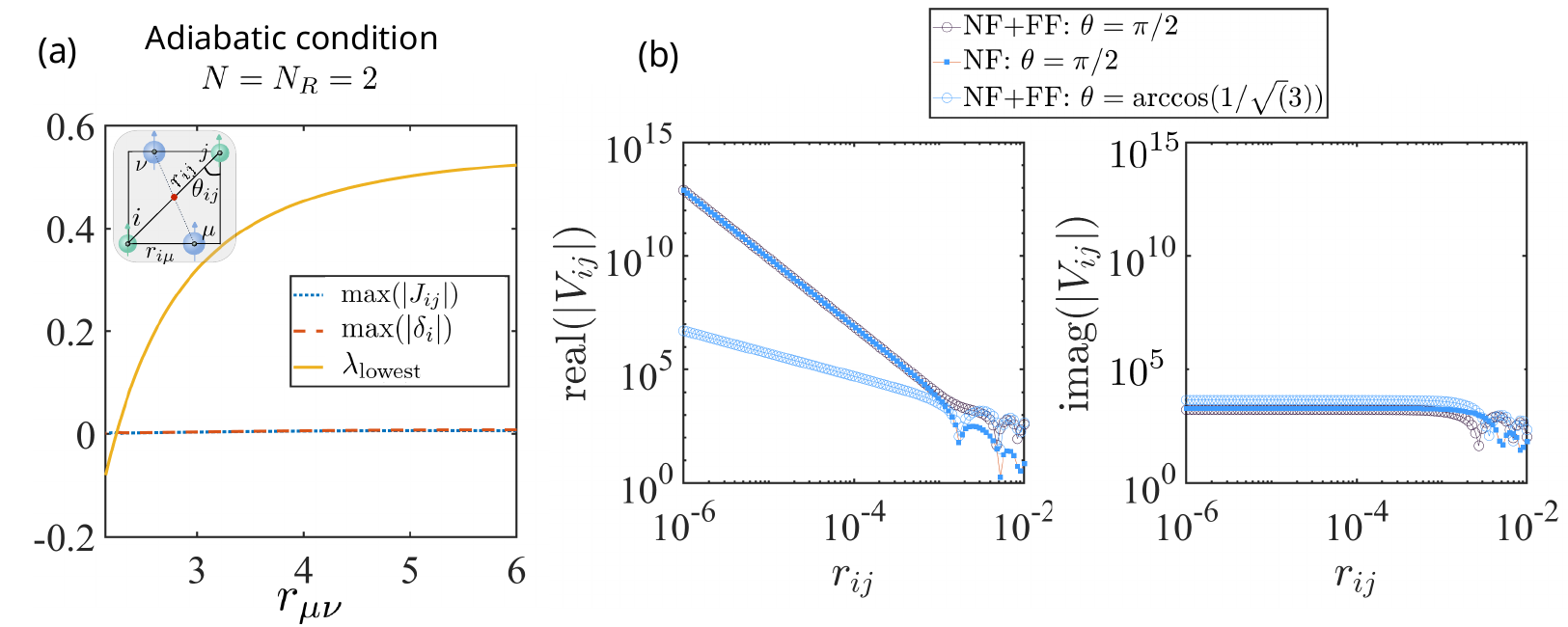}
    \caption{ (a) Adiabatic condition for two main and two relay atoms. We show the three parameters $\lambda_{\rm{lowest}} \gg \max(|\delta_i|), \max(|J_{ij}|)$ as a function of the distance between the relay atoms $r_{\mu \nu}$. For $r_{\mu \nu}>5$, the adiabatic condition is always valid. (b) Real (left) and imaginary (right) part of the near-field (NF) dipole-dipole couplings in blue and full dipole-dipole couplings (NF+FF) Eq.~\eqref{eq: fulldd} in red as a function of the distance $r_{ij}$ for $\theta = \pi/2$ and $\theta=\arccos(1/\sqrt(3))$. The case NF with $\theta=\arccos(1/\sqrt(3)) = 0$ and is not shown. We see that for $r_{ij} \ll 10^{-3}$ the FF term is always negligible compare to the NF terms.
    Parameters used for the simulation: $\gamma_i = 3800$ Hz, $\lambda = 7$ mm.}
    \label{fig: validity}
\end{figure}

The global condition for adiabatic elimination of the low-$\ell$ emitters requires that the eigenvalues $\lambda_i$ of the matrix $\bm{B} = -\mi\bm{M}$ (see main text for details on $\bm{M}$) dominate all other relevant energy scales in the system. This criterion has been rigorously established in~\cite{hagenmuller2020}. To systematically verify its validity, we compute the smallest eigenvalue, $\lambda_{\rm{lowest}} = \min(\lambda_i)$, and compare it against the maximal values of the relevant energy parameters, ensuring that $\lambda_{\rm{lowest}} \gg \max(\delta_i), \max(|J_{ij}|), \max(\gamma_i^{\rm{eff}})$.

To demonstrate this explicitly within our parameter regime, we consider a setup with two main and two relay atoms ($N = N_R = 2$). We fix the interatomic distances to $r_{ij} = 10$ and $r_{i\mu} = 8$, and vary the angle $\theta_{i\mu}$ to tune the distance between relay atoms, $r_{\mu\nu}$. Figure~\ref{fig: validity}(a) shows $\lambda_{\rm{lowest}}$ as a function of $r_{\mu \nu}$, confirming that it consistently exceeds both $\max(\delta_i)$ and $\max(|J_{ij}|)$.

These results demonstrate that the adiabatic elimination condition is well satisfied for all values $r_{\mu\nu} \gtrsim 5$ considered in this work. Moreover, in this regime, the dissipation rate $\gamma$ remains negligible throughout.

\section{Validity of the near-field regime}
\label{app: near_field}

An important aspect to verify is the validity of the near-field (NF) regime description used for the dipole-dipole interaction. Indeed, the full dipole-dipole interaction (NF+FF) is described by
\begin{equation}
\label{eq: fulldd}
\begin{array}{ll}
   V_{ij} = &- 3 \frac{\sqrt{\gamma_i\gamma_j}}{2} \left[\sin^2(\theta) \frac{\exp(i\xi)}{\xi} \right. \\
   & \left. + \big(3\cos^2(\theta)-1\big) \left(\frac{\exp(i\xi)}{\xi^3}
   - i \frac{\exp(i\xi)}{\xi^2}\right) \right]
\end{array}
\end{equation}
where $\xi = k r$, ($k=2\pi/\lambda$). Note that in our system, the transition frequencies are approximately 43 GHz, corresponding to a wavelength of $\lambda \approx 7 $ mm.

In Fig.~\ref{fig: validity}(b), we plot the real part (left) and imaginary part (right) of $|J_{ij}|$ from near-field description (see main text, blue) and Eq.~\eqref{eq: fulldd} (red) as functions of the distance $r_{ij}$ for two cases: $\theta_{ij} = \pi/2$ and $\theta_{ij}=\arccos(1/\sqrt3)$. For $\theta_{ij} = \pi/2$, the two equations start to diverge at $r_{ij} \approx 10^{-3} \; \rm{m} \approx \lambda$.

Importantly, the largest discrepancy occurs at the ``magic angle'' $\theta_{ij} = \arccos{(1/\sqrt{3})}$, where the near-field contribution cancels out entirely.  Even in this case, the value of $|J_{ij}|$ at short distances is several orders of magnitude smaller than for an arbitrary $\theta_{ij}$, suggesting minimal impact on our calculations. Additionally, note that the imaginary part of Eq.~\eqref{eq: fulldd} is several orders of magnitude smaller than the real part. This implies that induced dissipation can be safely neglected when $r_{ij} \ll 10^{-3}$.

\section{Comments on the mirrored configuration}
\label{app: symmetryconfig}

In this section, we explain why we primarily focus on a mirrored geometrical configuration for the relay atoms. This stems from the fact that the circular states must be at resonance to mediate an interaction and that the adiabatic elimination induces an effective detuning $\delta_i^{\rm eff}$ in each circular state. We found that the mirrored configuration ensures that the effective detuning is (nearly) identical for both circular states, facilitating the desired resonance condition. 

Let us demonstrate it in the case of two main atoms, denoted $A$, $C$ and two relay atoms, denoted $B_1$ and $B_2$. We have 
\begin{equation}
\begin{array}{cc}
    \bm{V}_{B_1} & = \begin{pmatrix}
                    V_{A, B_1} \\
                    V_{A, B_2}
    \end{pmatrix}, \; \; 
    \bm{V}_{B_2}   = \begin{pmatrix}
                    V_{C, B_1} \\
                    V_{C, B_2}
    \end{pmatrix}.
\end{array}
\end{equation}
If we have a central mirrored symmetry we know that
\begin{equation}
\begin{array}{lll}
    V_{A, B_1}  = & V_{C, B_2} =& V \\
    V_{A, B_2}  = & V_{C, B_1} =& \tilde{V}.\\
\end{array}
\end{equation}
Now, we define the matrix fo the internal dynamics of relay atoms by
\begin{equation}
    \bm{M}^{-1} = \begin{pmatrix}
                m_1 & m_2 \\
                m_3 & m_4
    \end{pmatrix}.
\end{equation}
One can show that to have $\delta^{\rm eff}_A = \delta^{\rm eff}_C$, the condition becomes $m_1 = m_4$ which is fulfilled in our case since $\bm{M}$ is defined as
\begin{equation}
    \bm{M} = \begin{pmatrix}
                \Delta & V_{B_1 B_2} \\
                V_{B_1B_2} & \Delta
    \end{pmatrix}.
\end{equation}
Then, using the invertible relation of a $2 \times 2$ matrix we see that $m_1 = m_4$. 
For larger system sizes, analytically demonstrating this condition becomes increasingly challenging. However, with some simplifications, we have verified the calculation for a system comprising three circular atoms and six relay atoms, showing that the effective detuning is equivalent for circular atoms located in the bulk. While the analytical approach is complex, verifying this condition numerically is straightforward. For instance, for $N = 21$, $N_R=40$, we obtain a difference between effective detuning of the order of $10^{-3} \; \rm GHz$ for $i=2\ldots 20$. Note that the two circular states at the edge ($i=1, i=21)$ have a detuning difference with the bulk of the order of $10^{-1} \; \rm GHz$.

\section{Comparison of Effective and Full Quantum Master Dynamics}
\label{app subsec: compa2sitesexact}

\begin{figure}[b]
    \centering
    \includegraphics[width=0.8\linewidth]{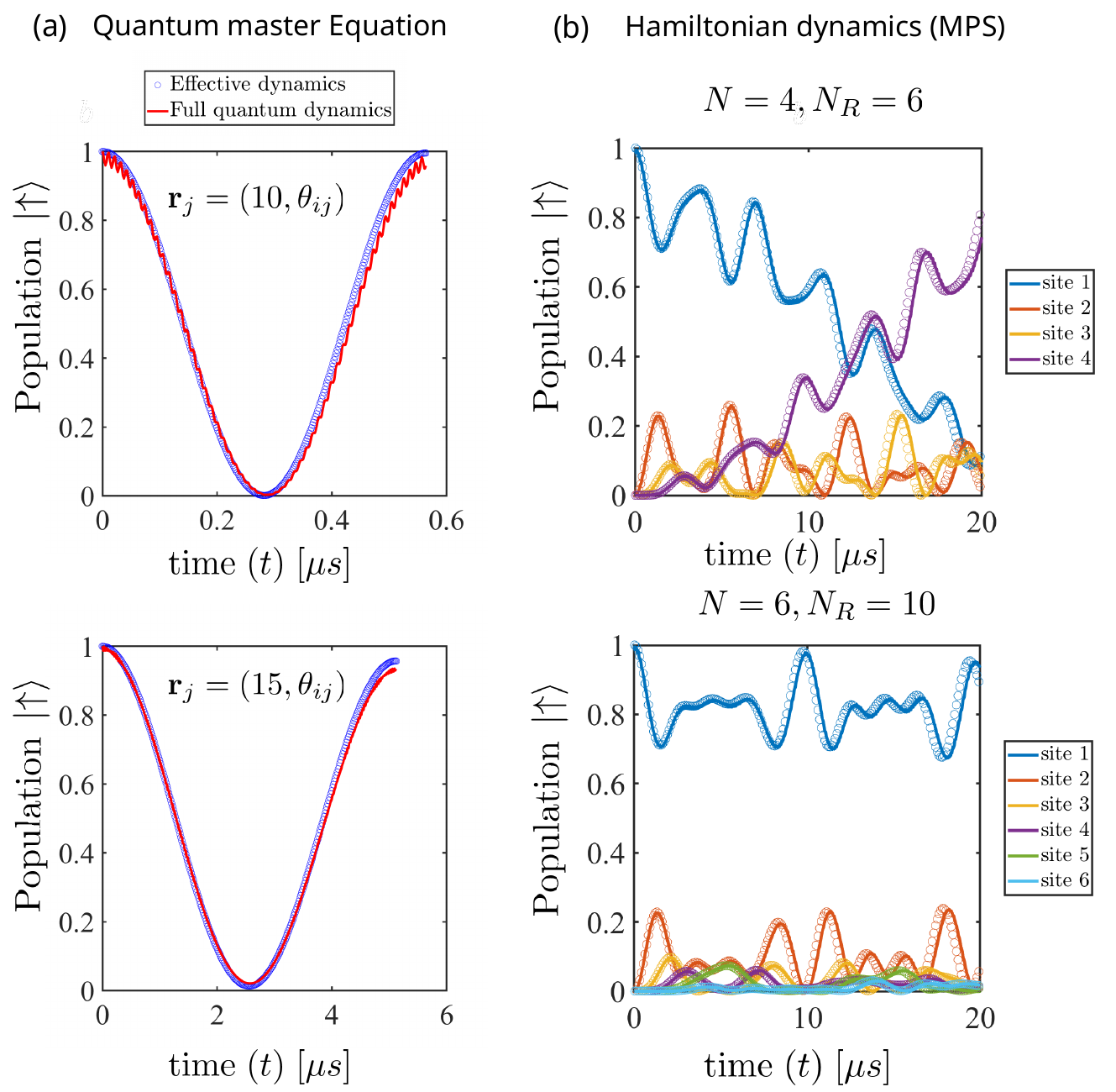}
    \caption{(a) Population in $\ket{\uparrow}$ at site $ i $ vs. time $ t $, computed from the full master equation (lines) and the effective model (dots) for two circular states. (b) MPS simulation of $\ket{\uparrow}$ population across the sites of the chain, comparing the effective (dots) and full (lines) Hamiltonian dynamics. [Parameters underlying the simulations are: $C_3^{(\uparrow\alpha)} = \pi \times 2.25$ GHz$\cdot\upmu$m$^3$,  $C_3^{(\beta\alpha)} = \pi \times 1.31$ GHz$\cdot\upmu$m$^3$,  $p_{\uparrow \downarrow}^{(2)} = -0.39$ MHz/(V/cm)$^2$,  $p_{\beta\alpha}^{(2)} = -57$ MHz/(V/cm)$^2$,  $F = 3.5$ V/cm, $ \textbf{r}_{i\mu} = (6, 0) $, $ r_{ij} = 15 $ (chain). Polar coordinates: $r_{ij}$ in $\upmu$m, $\theta_{ij}$ in rad. We also used the notation \textbf{r} = ($r_{ij}, \theta_{ij})$. For MPS simulation, we use a fourth order Trotter decomposition with a cutoff for the truncation error at $1E-13$, and a time steps $\delta_t = 0.01$.]}
    \label{fig: compa}
\end{figure}

In this section, we compare the effective dynamics with exact simulations of the full quantum system. We first focus on scenarios with two main and two relay atoms ($N = N_R = 2$), initializing the system at $t = 0$ with an excitation in $\ket{\uparrow}$ in one main atom. The evolution is computed using both the full quantum dynamics and the effective two-level system model [Eq.~\eqref{eq: effective_master_equation}]. Note that we consider two additional levels for the circular states to account for the BBR effects as explained in the main text.

Fig.~\ref{fig: compa}(a) shows the results over one oscillation period, where dots represent exact simulations and lines correspond to the effective model. We analyze two cases: $r_{ij} = 10\,\upmu\rm{m}$ (top) and $ r_{ij} =15\,\upmu\rm{m}$ (bottom) with $\theta_{ij} = \arccos(\sqrt{1/3})$, while keeping $r_{i\mu} = 6\,\upmu\rm{m}$, and $\theta_{i\mu} = 0$ fixed.
In Fig.~\ref{fig: compa}(b), we extend the comparison to pure Hamiltonian dynamics for a chain, using a Matrix Product State (MPS) simulation~\cite{schollwock2005the}. In particular, we use TEBD updates and a fourth-order Trotter decomposition with swap gates to implement long-range couplings. Here, we restrict ourselves to a two-level-system description since dissipation is not taken into account. Dots (plotted every 10 time steps for clarity) represent the effective simulation, while lines correspond to the full Hamiltonian dynamics. The top panel shows results for $N = 4$, and the lower panel for $N = 6$. The parameters are given in the caption.

In both cases, we find good agreement between the effective model and the full quantum dynamics.

\section{Analytical forms for the effective coupling matrix $J_{ij}$}

Following Eq.~\eqref{eq: effectivecoupling}, and neglecting interaction between relay atoms, the interaction between two main atoms can be written as 
\begin{equation}
J_{ij}  =  \frac{1}{\Delta}\sum_{\mu} V_{\uparrow\alpha}(r_{i\mu}, \theta_{i\mu}) V_{\uparrow\alpha}(r_{\mu j}, \theta_{\mu j}) \ .
\end{equation}
This equation ($N=N_R=2$) is simplified when the position vector $\mathbf{r}_\mu = (r_{i\mu}, \beta_{i\mu} = \theta_{i\mu} - \pi/2)$ is fixed and we consider a mirrored configuration as described in the main text. In this case, the solution can be expressed in terms of a reduced set of parameters (depending solely on indices $i\mu$ and $ij$ due to the mirrored configuration). We then have 
\begin{equation}
V_{\uparrow\alpha}(r_{\mu j}, \theta_{\mu j}) = f(r_{ij}, r_{i\mu}, \theta_{i\mu}, \theta_{ij}),
\end{equation}
which naturally follows from the symmetry of the configuration. Specifically, one finds that  
\begin{equation}
\begin{array}{ccl}
\label{eq: analyticalscaling}
J_{ij}  =  \frac{1}{\Delta} \sum_{\mu} V_{\uparrow\alpha}(r_{i\mu}, \theta_{i\mu}) V_{\uparrow\alpha}(r_{\mu j}, \theta_{\mu j})
 = \frac{2}{\Delta}V_{\uparrow \alpha}(r_{i\mu}, \theta_{i\mu})  
 \Bigg[ \frac{1}{2r_{ij}^3} \Big( \frac{\frac{B}{r_{ij}} + \frac{C}{r_{ij}^2}}{(1 + \frac{D}{r_{ij}} + \frac{E}{r_{ij}^2} )^{5/2}} \Big) \Bigg],
\end{array}
\end{equation}
where $B = -2r_{i\mu} \cos(\beta_{i\mu} - \beta_{ij}) + 6r_{i\mu} \cos(\beta_{i\mu} + \beta_{ij})$, $C = -3r_{i\mu}^2 \cos(2\beta_{i\mu}) + r_{i\mu}^2$, $D = -2r_{i\mu} \cos(\beta_{i\mu} - \beta_{ij})$, and $E =  r_{i\mu}^2$.
From this, one can perform a Taylor series expansion around the expansion point $1/r_{ij}$ in order to obtain Eq.~\eqref{eq: analyticalscaling_taylor}, with details up to the seventh order we obtain
\begin{equation}
\begin{array}{ccl}
\label{eq: analyticalscaling_taylor}
J_{ij}  =  \frac{2}{\Delta}V_{\uparrow\alpha}(r_{i\mu}, \theta_{i\mu}) \Bigg[ 
  \frac{B}{2r_{ij}^4} + \frac{C/2 - 5 B D/4}{r_{ij}^5}  - \frac{\left(\frac{B}{2} \left(\frac{5E}{2}
 - \frac{35D^2}{8} \right) + \frac{5CD}{4} \right)}{r_{ij}^6} \Bigg] + \mathcal{O}(r_{ij^{-7}}).
\end{array}
\end{equation} 

\section{Quality of fits and exclusion around magic angle line}

\begin{figure}[htb]
    \centering
    \includegraphics[width=1\linewidth]{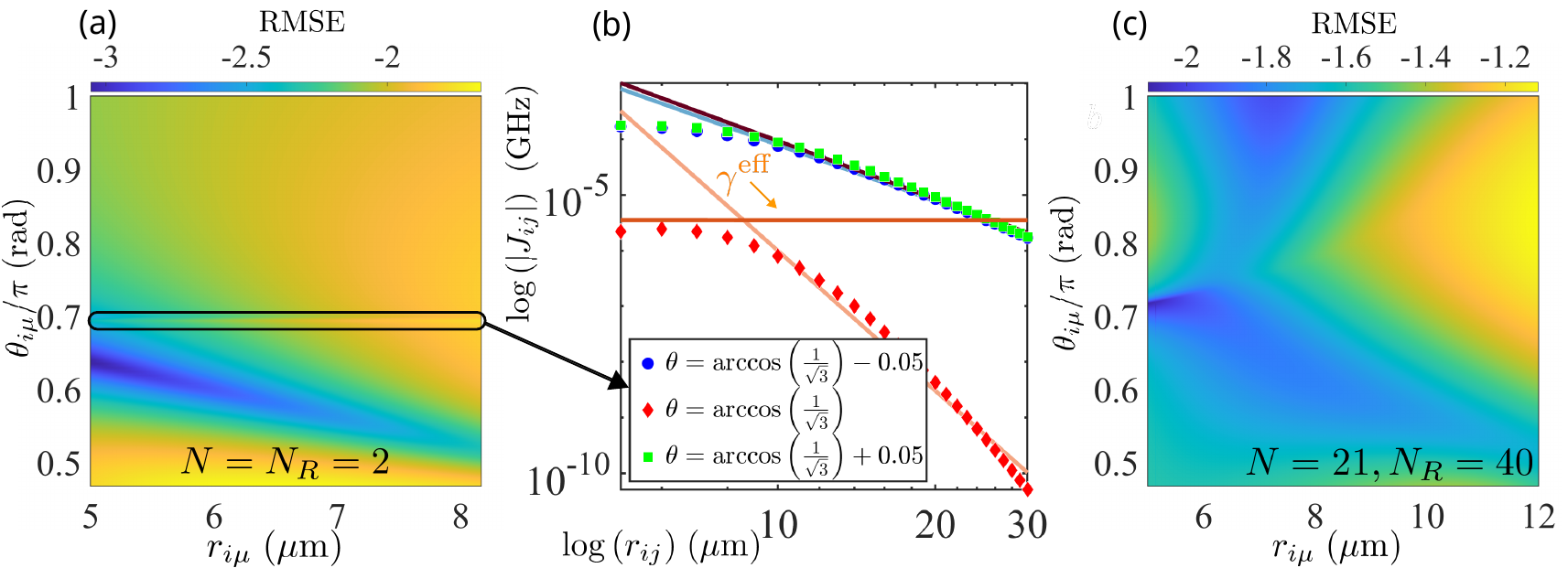}
    \caption{(a),(c) Contour plots of the root mean square errors from the fitting procedure in Fig.~\ref{fig: fig2} as a function of $ r_{i\mu} $ and $\theta_{i\mu}.$ for $N=N_R=2$ (a) and a chain with $N=21, N_R=40$ (c). (b) Examples of the distance-scaling of $ |J_{ij}| $ to highlight the exclusion area around the magic angle. Parameters  identical to Fig.~\ref{fig: model}.}
    \label{fig: qualityfitfig2}
\end{figure}

In this section, we assess the quality of the fits shown in Fig.~\ref{fig: fig2} and discuss an exclusion area near the magic angle line for the case $ N = N_R = 2 $. The root mean square errors (RMSE) from the fitting procedure, used to extract the exponent $ b $ (see Fig.~\ref{fig: model} and Fig.~\ref{fig: fig2}), are presented in Fig.~\ref{fig: qualityfitfig2}(a) and (c) for $ N = 2 $ and $ N = 21 $, respectively. A distinct line near the magic angle in Fig.~\ref{fig: fig2}(a) and Fig.~\ref{fig: qualityfitfig2}(a) shows sharp deviations in both the exponent and RMSE from other regions. This behavior is due to the cancellation of near-field dipole interactions, making the extraction of exponent $ b $ meaningless in this vicinity, as illustrated in Fig.~\ref{fig: qualityfitfig2}(b).

To further address this issue, we define an exclusion region around the magic angle configuration using the criterion $ |J_{ij}| < \gamma^{\rm eff} $ for $ r_{ij} \leq 25 \, \upmu\text{m} $. The boundaries of this exclusion are shown in Fig.~\ref{fig: qualityfitfig2}(b).

\end{widetext}

\end{document}